\documentclass[conference]{IEEEtran}
\IEEEoverridecommandlockouts
\usepackage{cite}
\usepackage{amsmath,amssymb,amsfonts}
\usepackage{autobreak}
\usepackage{algorithmic}
\usepackage{textcomp}
\usepackage{xcolor}
\usepackage{url}
\usepackage{algorithmic}
\usepackage{algorithm}
\usepackage{braket}
\usepackage{mathtools}
\usepackage{subfigure}
\usepackage{here}

\def\BibTeX{{\rm B\kern-.05em{\sc i\kern-.025em b}\kern-.08em
    T\kern-.1667em\lower.7ex\hbox{E}\kern-.125emX}}
\begin{document}
\title{Parametrized Energy-Efficient Quantum Kernels for Network Service Fault Diagnosis\\
}
\makeatletter
\newcommand{\linebreakand}{%
  \end{@IEEEauthorhalign}
  \hfill\mbox{}\par
  \mbox{}\hfill\begin{@IEEEauthorhalign}
}
\makeatother

\author{
\IEEEauthorblockN{Hiroshi Yamauchi}
\IEEEauthorblockA{\textit{Research Institute of Advanced Technology} \\
\textit{SoftBank Corp.}\\
hiroshi.yamauchi@g.softbank.co.jp}

\and
\IEEEauthorblockN{Tomah Sogabe}
\IEEEauthorblockA{
\textit{The University of Electro-Communications}\\
sogabe@uec.ac.jp}

\linebreakand
\IEEEauthorblockN{Rodney Van Meter}
\IEEEauthorblockA{
\textit{Keio University}\\
rdv@sfc.wide.ad.jp}
}
\maketitle

\begin{abstract}

In quantum kernel learning, the primary method involves using a quantum computer to calculate the inner product between feature vectors, thereby obtaining a Gram matrix used as a kernel in machine learning models such as support vector machines (SVMs). 
However, a method for consistently achieving high performance has not been established.
In this study, we investigate the diagnostic accuracy using a commercial dataset of a network service fault diagnosis system used by telecommunications carriers, focusing on quantum kernel learning, and propose a method to stably achieve high performance.
We show significant performance improvements and an efficient achievement of high performance over conventional methods can be attained by applying quantum entanglement in the portion of the general quantum circuit used to create the quantum kernel, through input data parameter mapping and parameter tuning related to relative phase angles.
Furthermore, experimental validation of the quantum kernel was conducted using IBM's superconducting quantum computer IBM-Kawasaki, and its practicality was verified while applying the error suppression feature of Q-CTRL's Fire Opal.
\end{abstract}

\begin{IEEEkeywords}
quantum computer, quantum machine learning, quantum artificial intelligence, quantum support vector machines, quantum kernel learning, network system, fault diagnosis
\end{IEEEkeywords}

\section{Introduction}

\begin{figure*}[t!]
\begin{center}
\includegraphics[width=139mm]{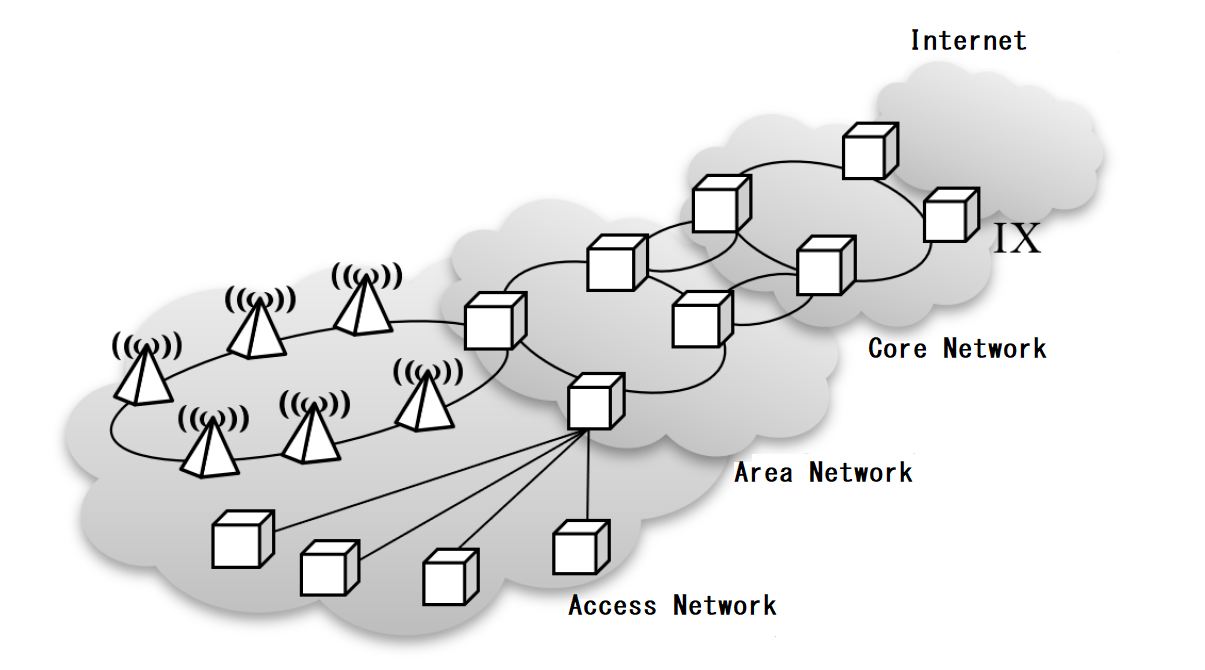}
\caption{{\small Nation-wide Telecommunication Network}}
\label{fig:nw}
\end{center}
\end{figure*}

Quantum kernel learning is an effective algorithm as a machine learning method utilizing quantum computers. 
The capability for super spatio-temporal dimensional feature representation in Hilbert space, leveraging quantum mechanical principles, is considered to have promising advantages in learning efficiency over classical computers.
Recently, research on quantum AI (Artificial Intelligence) in industries such as health-care, finance, and manufacturing has also become increasingly active\cite{bib:qml1}\cite{bib:qml2}\cite{bib:qml3}.

The similarity between quantum computing and kernel methods in machine learning suggests efficient computation in a vast Hilbert space through the process of encoding input data into quantum states\cite{bib:qml}.
Quantum kernel learning involves the use of kernels calculated by the quantum computers to build inference models using support vector machines\cite{bib:svm}, taking advantage of the quantum computer's ability to extract features in an ultra-high-dimensional space-temporal space through quantum mechanics.

This is especially an effective approach within the current NISQ (Noisy Intermediate-Scale Quantum device) systems\cite{bib:nisq}, making it an important technology in practical applications.
In recent studies of quantum-classical hybrid computers, many investigations have been conducted on various quantum kernels. 
The impact of circuit depth and expressivity through variational optimization methods\cite{bib:emb} and the identification of promising classes for specific datasets with group theoretic structures\cite{bib:cov} are among the topics of particular interest in recent studies on quantum kernel learning using variational optimization.
In this study, we focused on deriving optimal inference performance by using quantum kernel circuits for the measurement of expectation values in inner product calculations without employing variational optimization. 

Research involving use cases related to industrial applications is actively being presented by various companies.
The effectiveness of image classification using quantum kernels on large datasets has been demonstrated for cloud detection in multispectral satellite imagery within satellite data analysis\cite{bib:sat}.
Studies have been conducted on the automated segmentation and classification of hand thermal images in rheumatoid arthritis, comparing the use of machine learning algorithms with quantum machine learning techniques\cite{bib:rheu}. 

These studies report that the results of quantum algorithms are comparable to those of classical algorithms.
Theoretical studies have shown provable advantages on synthetic data sets, but it has not been clarified whether quantum advantages are attainable and with which type of data sets\cite{bib:powerkernel}. 
In one study using electronic health record data, the authors also report examples investigating how the accuracy of a model contributes in relation to the number of features and sample size\cite{bib:ehr}.

In the field of telecommunications, the implementation of 5G and beyond networks brings about ultra-fast, high-capacity, and low-latency information transmission technologies, along with the expansion of device-to-device communication such as IoT sensor broadcast data and high-definition video streaming data. 
In the future networks, there is an increasing need to efficiently handle huge volumes of data, along with a demand for network functionality scalability that incorporates AI learning capabilities \cite{bib:airan}. Additionally, there is an expectation for accelerated computational capabilities through quantum computers, serving as the backend's high-performance computing infrastructure. 

The configuration of the target commercial network is comprised of multiple-layered networks nation-wide as shown in {\mdseries Fig. \ref{fig:nw}}, with the highest level connected to the Internet via an Internet Exchange (IX), a core network that connects major cities like arteries, area and access networks that are deployed across regions and mobile base station and optical fiber fixed access lines that cover local bases like capillaries.

Within such a network configuration, a vast variety and number of devices that accommodate communication services are installed, and operation engineers carry out missions such as fault diagnosis and recovery in operating these devices using various commands to ensure service integrity.
As the expansion of communication demand in recent years, network scales have grown, and the complexity of diagnostic schemes has increased, prompting automation within systems, especially research into approaches using machine learning \cite{bib:darkstar} \cite{bib:proactive} \cite{bib:fog} \cite{bib:trios}. 

In this study, we have conducted a study on using quantum kernels for fault diagnosis tasks in a commercial telecommunication network.

This study explored a method that implements tunable parameters in the quantum entanglement generation part of quantum kernels, which allows for more stable extraction of better performance in terms of inference performance of quantum machine learning. 
Performance verification of the proposed method was conducted using the IBM superconducting quantum computer IBM-Kawasaki, and the practical level of performance was confirmed using Q-CTRL's Fire Opal\cite{bib:qctrl}, an effective error suppression product for NISQ machines.
In the context of quantum computer hardware, we will also review the fact that machine learning performance using quantum kernels is approaching a practically viable level at the current time.

\section{Methods}

\begin{figure}[t!]
\begin{center}
\includegraphics[width=80mm]{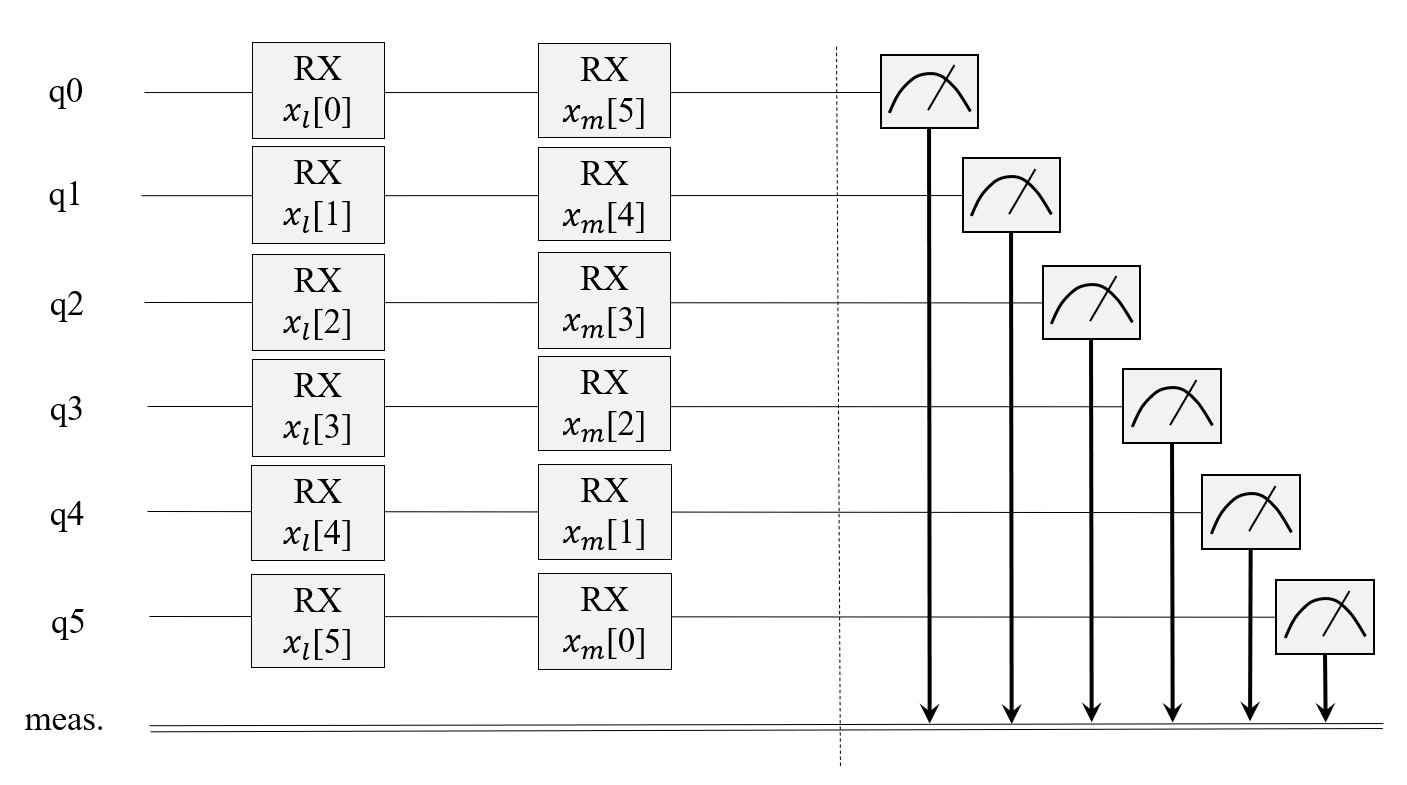}
\caption{{\small Conventional quantum kernel circuit. The configuration involves parameterizing the two feature vectors $\vec{x_l},\vec{x_m}$ targeted for inner product calculation with Pauli gates.}}
\label{fig:circ1}
  \end{center}
\end{figure}

\begin{figure*}[t!]
\begin{center}
  \subfigure[Entanglement generation part]{
   \includegraphics[width=125mm]{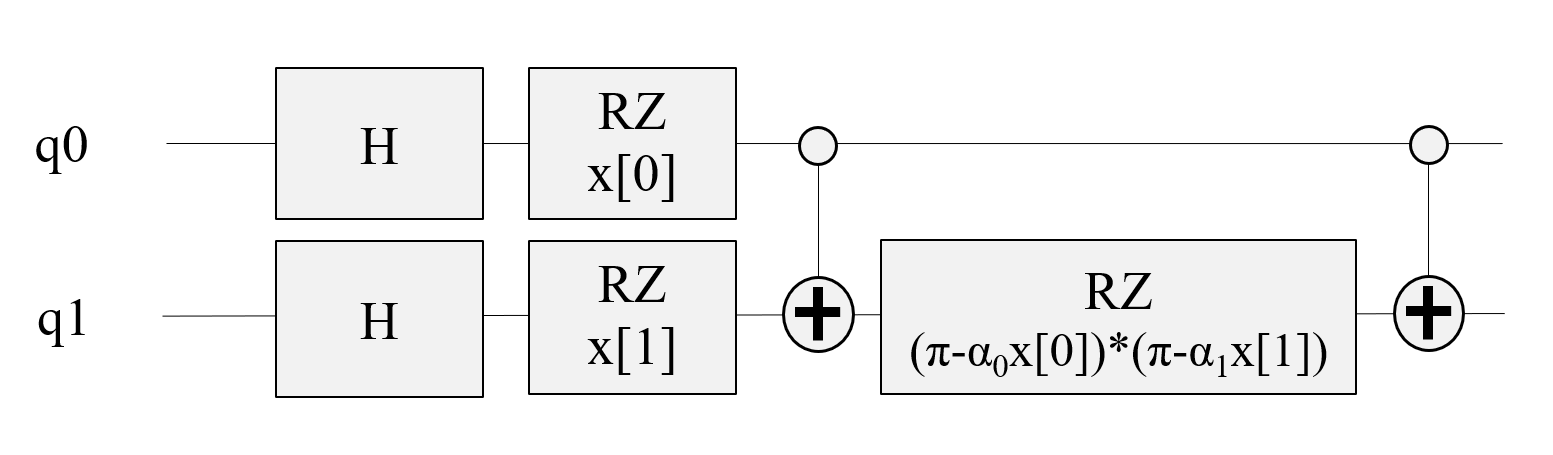}
  }\\
  \subfigure[Overall circuit]{
   \includegraphics[width=175mm]{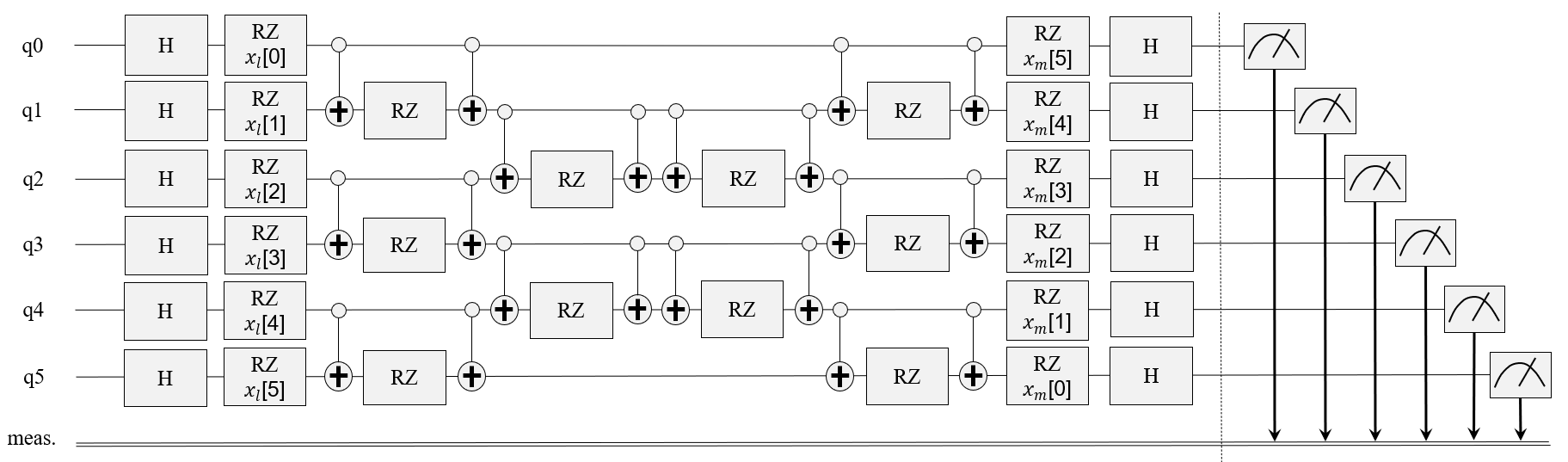}
  }
\caption{{\small Parametrized Energy-Efficient Quantum Kernels. For two qubits parameterized with Pauli gates, the phase amplitude further parameterized by Z-rotation gates, as shown in the entanglement generation section, can be tuned in the form represented by Equation (4) for global parameters.}}
\label{fig:circ2}
\end{center}
\end{figure*}

\begin{figure*}[t!]
\begin{center}
\includegraphics[width=120mm]{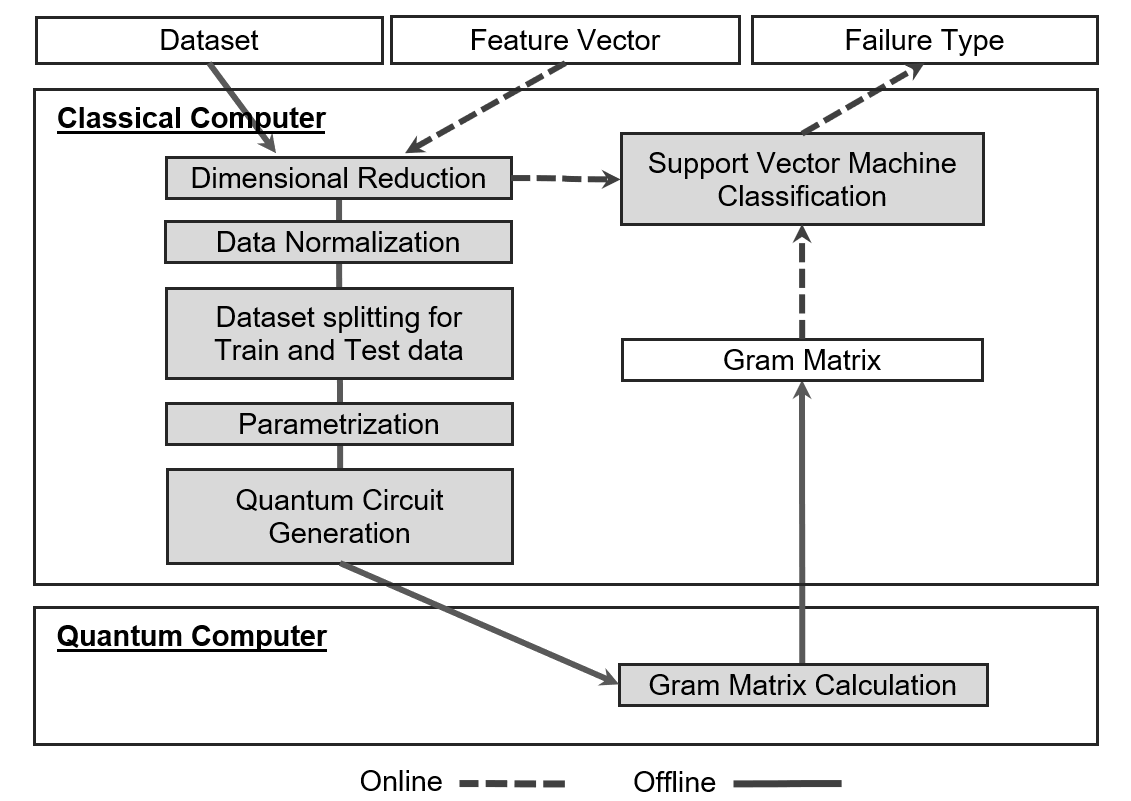}
\end{center}
\caption{{\small Network Service Fault Diagnosis System Diagram. We conduct training using a dataset extracted from the offline diagnostic system logs over a fixed period, and use the Gram matrix generated by the quantum kernel for online service fault type estimation via support vector machines. The system is configured as a hybrid of classical and quantum components, where we intend to effectively utilize feature representations generated by quantum computers.}}
\label{fig:systemdiagram}
\end{figure*}

In quantum kernel learning, the method for creating kernels is obtained by calculating the inner product between feature vectors $\vec{x_l}$, $\vec{x_m}$ of the training dataset on a quantum computer as follows:
\begin{equation}
k(\vec{x_l},\vec{x_m})=|\braket{\phi({\vec{x_l}})|\phi(\vec{x_m})}|^2 \label{eq1}
\end{equation}
where $l,m$ represent the pattern index numbers in the dataset.

For the initial quantum circuit state, gate operators $U_{\phi(x)}$ are used to obtain $\phi(x)$ for parameter mapping of quantum features on unitary spaces.
\begin{equation}
k(\vec{x_l},\vec{x_m})=|\bra{0}^{{\otimes}n} U^{\dagger}_{\phi(\vec{x_l})} U_{\phi(\vec{x_m})} \ket{0}^{{\otimes}n} |^2 \label{eq2}
\end{equation}
where $n$ represents the number of qubits, which is the dimensionality of the parameterized feature vector.

As a conventional method,  {\mdseries Fig. \ref{fig:circ1}} illustrates a technique for computing a Gram matrix using a quantum circuit that employs the number of qubits based on the feature dimension number, with parametrization performed on Pauli gates.

In our proposed method, the quantum circuit used for the quantum kernel learning, ``Parametrized Energy-Efficient Quantum Kernels" is shown in {\mdseries Fig. \ref{fig:circ2}}. 
In the parameterization of the two qubit gate, implementing tunable parameter settings that allow for efficient control of entanglement strength through complex parameterization of phase rotation in the Z-rotation enables the system to effectively utilize higher-dimensional feature spaces for improved performance.
To emphasize the benefits of unitarity, which is crucial in quantum computing, the input data is normalized and then parameterized onto qubits.
The gate operator $U_{\phi(x)}$ in our proposed method is

\begin{multline}
U_{\phi(\vec{x})}=\\
(\exp(i\Sigma_{p,q}\phi_{p,q}(\vec{x})Z_{p}\otimes Z_{q})
\exp(i\Sigma_{p}\phi_{p}(\vec{x})Z_{p})H^{\otimes n})^{d}\label{eq3}
\end{multline}
where p, q are the index numbers of adjacent qubits and $d$ represents the circuit component depth.

The parameter mapping to Z-rotation gate of the quantum entanglement generation part between two qubits is
\begin{equation}
\phi_{p,q}(\vec{x})=(\pi-\alpha_{p} \vec{x_{p}})*(\pi-\alpha_{q} \vec{x_{q}})\label{eq5}
\end{equation}
where the values of the coefficients $\alpha_{p}$ and $\alpha_{q}$ can be tuned.

In subsequent evaluations, $\alpha_{p}$ is set equal to $\alpha_{q}$.
It should be noted that the case of $\alpha=1$ is commonly used and known as the ZZFeatureMap\cite{bib:ibm}.

The result of inner product calculations using quantum circuits can be obtained by measuring the value of $\ket{0}^{{\otimes}n}$ in the quasi-probability distribution of the circuit's output. Gram matrix $K$ is calculated by mapping the inner product values of each feature vector as follows, 

\begin{equation}
K=
\begin{pmatrix} 
  k(\vec{x_{1}},\vec{x_{1}}) & k(\vec{x_{1}},\vec{x_{2}}) & \dots  & k(\vec{x_{1}},\vec{x_{m}}) \\
  k(\vec{x_{2}},\vec{x_{1}}) & k(\vec{x_{2}},\vec{x_{2}}) & \dots  & k(\vec{x_{2}},\vec{x_{m}}) \\
  \vdots & \vdots & \ddots & \vdots \\
  k(\vec{x_{l}},\vec{x_{1}}) & k(\vec{x_{l}},\vec{x_{2}}) & \dots  & k(\vec{x_{l}},\vec{x_{m}})
\end{pmatrix}
\end{equation}

The Gram matrix $K$ is used by the support vector machine classifier of the kernel function to project a high-dimensional space to generalize a nonlinearly separable data\cite{bib:svm2}. Formally if we have the data $\vec{x_l},\vec{x_m} \in X$ and a map $\phi : X \implies \mathbb{R}^n$ then

\begin{equation}
K(\vec{x_l},\vec{x_m})=\langle \phi(\vec{x_l}), \phi(\vec{x_m})\rangle.
\end{equation}

As mentioned above, using quantum kernels, we employ the learned Gram matrix from the dataset to implement the diagnostic function on the system, as shown in {\mdseries Fig. \ref{fig:systemdiagram}}.

\section{Experiments}

In this section, we describe the dataset used in the experimental evaluation of the proposed method, incorporating the problem setting in the context of the use case background. 
Next, we conduct comparative experiments between classical and quantum algorithms, and finally, we review experiments conducted on quantum computer hardware using the proposed method.

\subsection{Dataset of network service fault diagnosis}

\begin{figure*}[t!]
\begin{minipage}{.50\textwidth}
\begin{center}
\includegraphics[width=90mm]{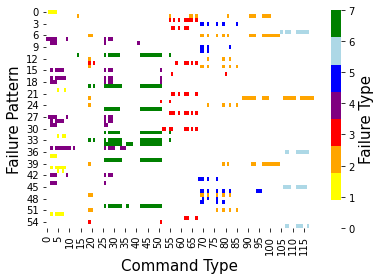}
\end{center}
\caption{\small Network service fault diagnosis dataset. 
The horizontal axis represents the 120 types of network device commands used for fault diagnosis, the vertical axis shows unique series indicating fault type, and the color of each plot indicates the type of fault as shown in Table I. The plot indicates cases where the diagnostic results were abnormal for each command.}
\label{fig:dataset}
\end{minipage}
\begin{minipage}{.50\textwidth}
\begin{center}
\makeatletter
\def\@captype{table}
\makeatother
\caption{Failure Type}
\begin{tabular}{clc}
\hline
ID & Failure Type\\
\hline
1 & Error detection at edge router\\
2 & Optical power failure at terminal on user site\\
3 & Optical signal degradation on subscriber segment\\
4 & Configuration error at equipment on station\\
5 & Interface error on central relay section\\
6 & Packet error on central relay section\\
7 & Device error at equipment on station\\
\hline
\end{tabular}
    \end{center}
\label{table:label}
\end{minipage}
\end{figure*}

This study evaluates the performance of quantum kernels using a dataset for fault diagnosis extracted from logs of a commercial system as known as ``TRIOS" of in-house developed network operation system for our network service of SmartVPN/SmartInternet \cite{bib:svpn}. 
The dataset is constructed based on investigation logs of service failure reports over three months.
 {\mdseries Fig. \ref{fig:dataset}} is a dataset extracted from system logs for fault diagnosis.
It shows a structure where one access network inter-connects with three types of core networks, providing service paths to each user site.
It consists of 56 failure patterns featuring normal and abnormal results from 120 types of commands as characteristics, with 7 types of fault categories.
Failure types are defined as shown in Table I, serve as information for diagnosing where in the communication segment the cause of a network service failure is located, enabling swift restoration. These definitions are used as the supervised labels for the dataset.

\subsection{Classification performance evaluation of Parametrized Energy-Efficient Quantum Kernel by Tensor network simulation}

\begin{figure*}[t!]
\begin{center}
\includegraphics[width=140mm]{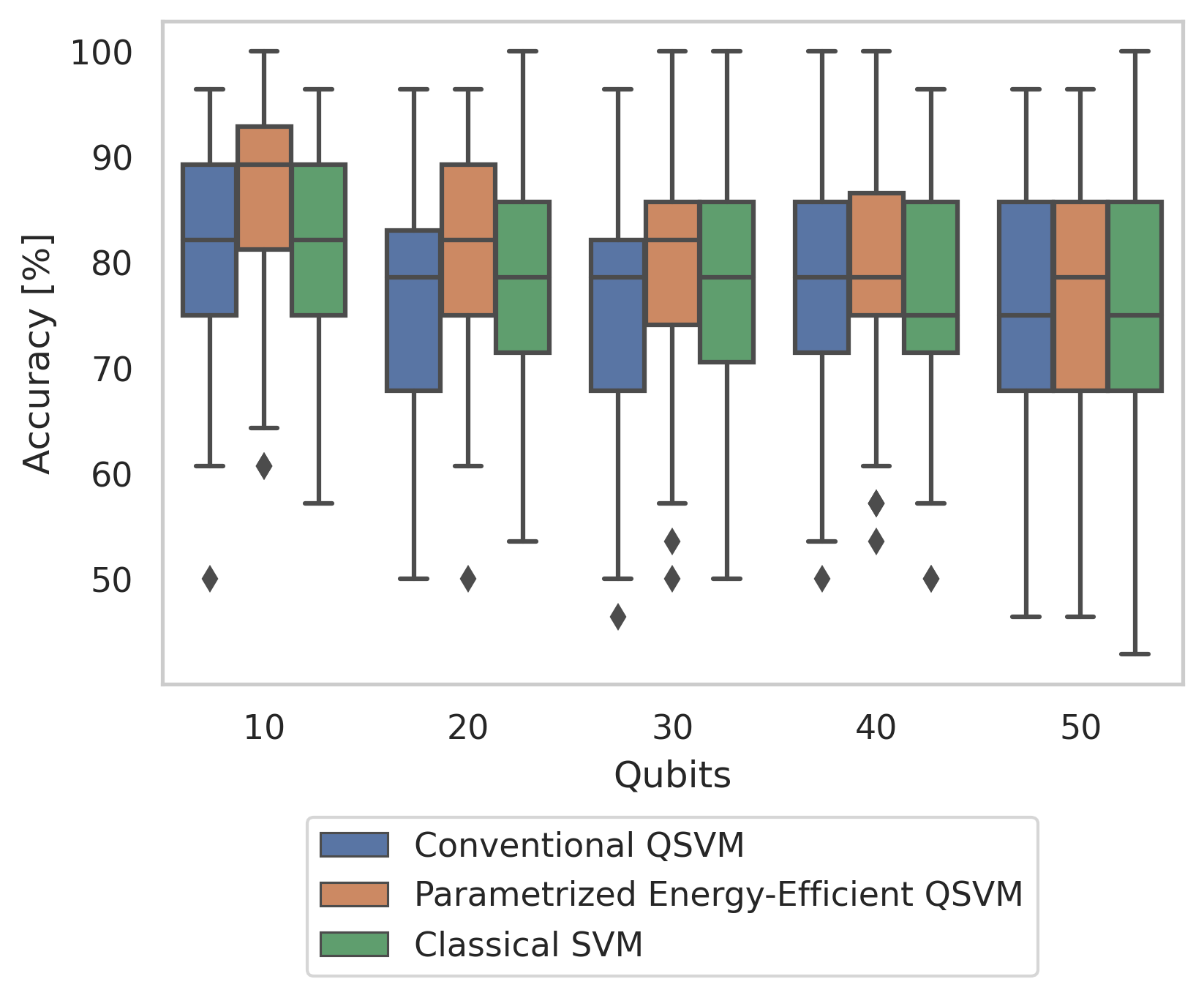}
\end{center}
\caption{{\small Classification performance between conventional and proposed QSVM by Tensor network simulation comparing to classical SVM. Each box-and-whisker plot shows the distribution of the classification accuracy for each qubits in 100 trials of crossvalidation split pattern. The box shows the quartiles of the data while the whiskers extend to show the rest of the distribution and diamond is outlier in the 95\% confidence interval.}}
\label{fig:figrand}
\end{figure*}

\begin{figure*}[t!]
\begin{center}
  \subfigure[10 Qubits]{
   \includegraphics[width=55mm]{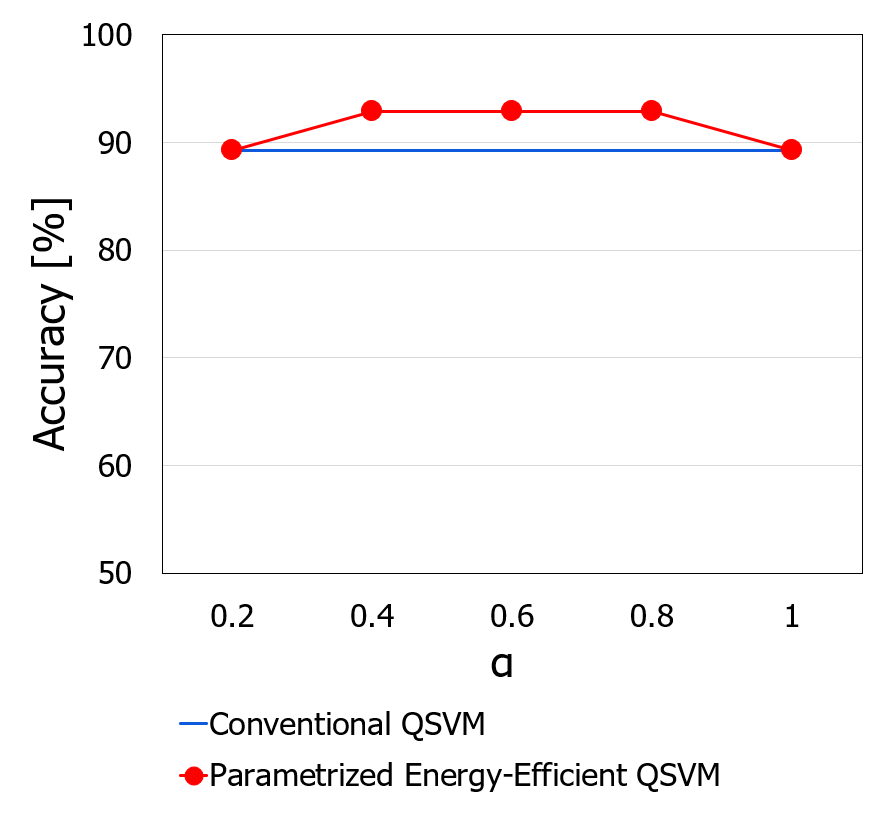}
  }~
  \subfigure[20 Qubits]{
   \includegraphics[width=55mm]{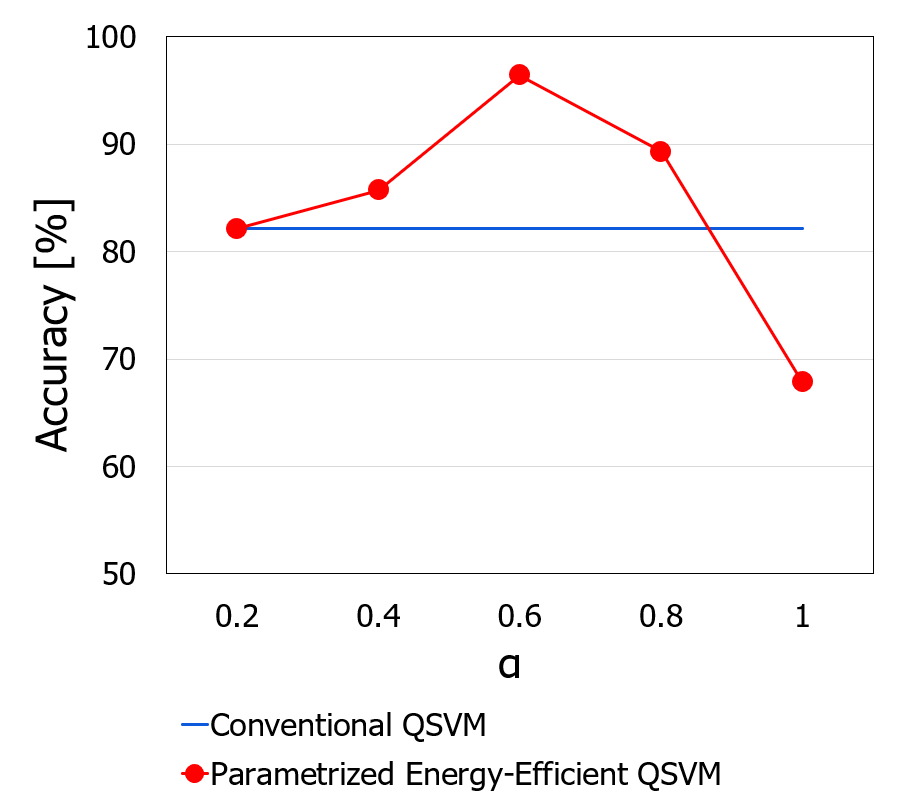}
  }~
  \subfigure[30 Qubits]{
   \includegraphics[width=55mm]{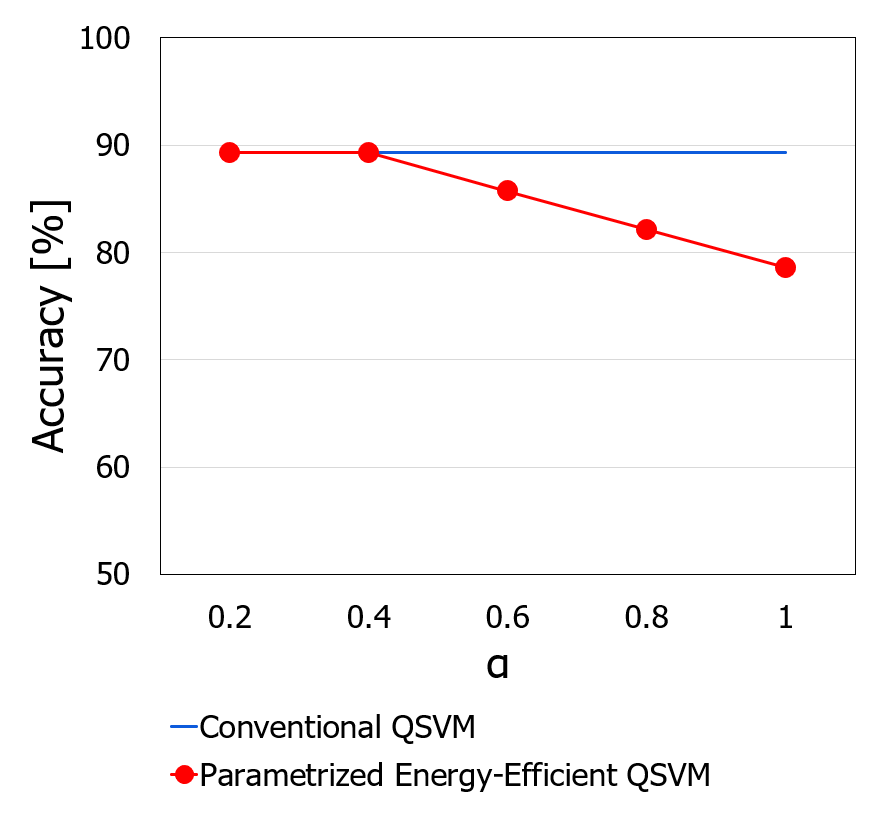}
  }
\caption{{\small $\alpha$ parameter dependency of classification accuracy by proposed circuit for each qubits.
In each qubit, the values indicated by the blue lines represent those of the conventional method, while the red plots show the transition of classification accuracy for each $\alpha$ parameter setting in the proposed method.}} 
\label{fig:alphaparam}
\end{center}
\end{figure*}

To evaluate this proposed method, we preprocess the dataset according to the following procedure. First, we perform dimensionality reduction to match the number of parameters with the number of qubits in the input quantum circuit, and then normalize these parameters before parameterizing them into the qubits. In this case, PCA (Principal Component Analysis) was used for the dimensionality reduction algorithm.

The kernel is created by a quantum kernel circuit parametrized for the feature vectors $\vec{x_l}, \vec{x_m}$ of training data and computing the inner product to obtain the Gram matrix for all combinations of training data.
This kernel is then used in a SVM to evaluate the classification performance on the test data.

{\mdseries Fig. \ref{fig:figrand}} shows a comparison of the classification performance of a conventional QSVM (Quantum Support Vector Machine), our proposed Parametrized Energy-Efficient QSVM using a tensor network simulator\cite{bib:quimb}\cite{bib:tn} and classical SVM using the RBF (Radial Basis Function) kernel for register sizes of 10 to 50 qubits.
The box plot data for each model shows the distribution of classification accuracy based on 100 split patterns for 50\% cross-validation.
The proposed method demonstrates superior performance compared to conventional QSVM, comparing the overall average values of the evaluation results for all qubit numbers in terms of accuracy, the values were 77\%, 81\%, and 78\% for conventional QSVM, Parametrized Energy-Efficient QSVM, and Classical SVM, respectively.

In subsequent evaluations, the QSVM was evaluated using a single split sample data for each number of qubits, where the classification accuracy of the classical SVM ranged from 85\% to 89\%.

The classification performance evaluated while tuning the $\alpha$ parameter of Parametrized Energy-Efficient quantum kernels for each register size is shown in {\mdseries Fig. \ref{fig:alphaparam}}. It can be seen that the performance of the proposed QSVM exceeds that of the conventional QSVM at a specific value of $\alpha$ parameter. 
By increasing the number of qubits, optimal inference performance 96\% is achieved at 20 qubits and the optimal setting for the $\alpha$ parameter was found to be between 0.4 to 0.6.

It has been shown that superior performance in quantum kernel learning can be efficiently achieved by adjusting the parametrized phase intensity using the $\alpha$ parameter with the proposed method. We believe that the optimal number of qubits and the $\alpha$ parameter are determined by the size of the dataset, and operations considering this are assumed in use cases.

\subsection{Hardware experiment using IBM's superconducting quantum computer and performance improvement by Q-CTRL's Fire Opal error suppression}

\begin{figure}[t!]
\begin{center}
\includegraphics[width=70mm]{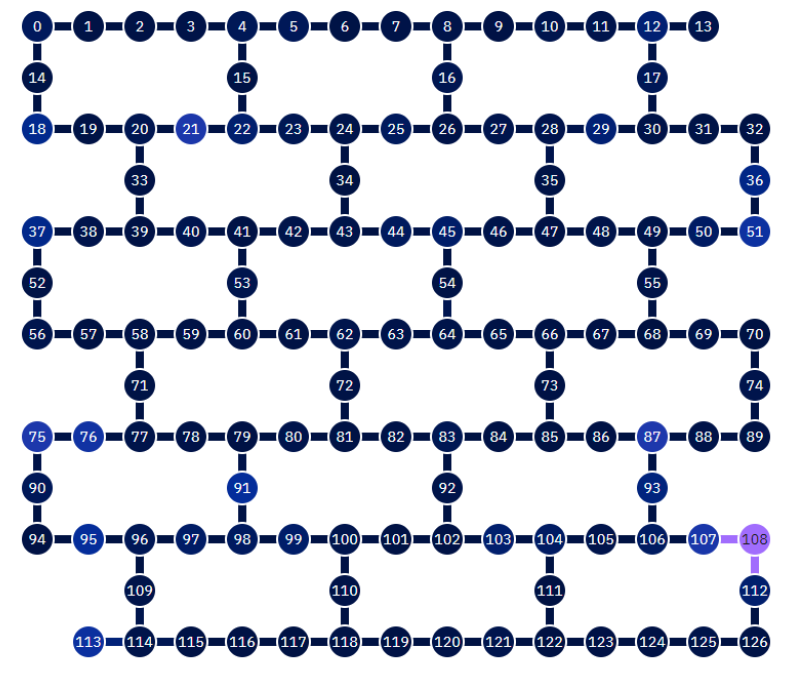}
\end{center}
\caption{{\small QPU architecture of 127-qubit IBM-Kawasaki. Processor type: Eagle r3, EPLG (Error per layered gate for a 100-qubit chain): 2.4\%, Median ECR error: 7.470e-3, Median SX error: 2.252e-4, Median readout error: 1.090e-2, Median T1: 194.33 $us$, Median T2: 146.9 $us$}}
\label{fig:qpu}
\end{figure}

\begin{figure*}[t]
\begin{center}
\includegraphics[width=130mm]{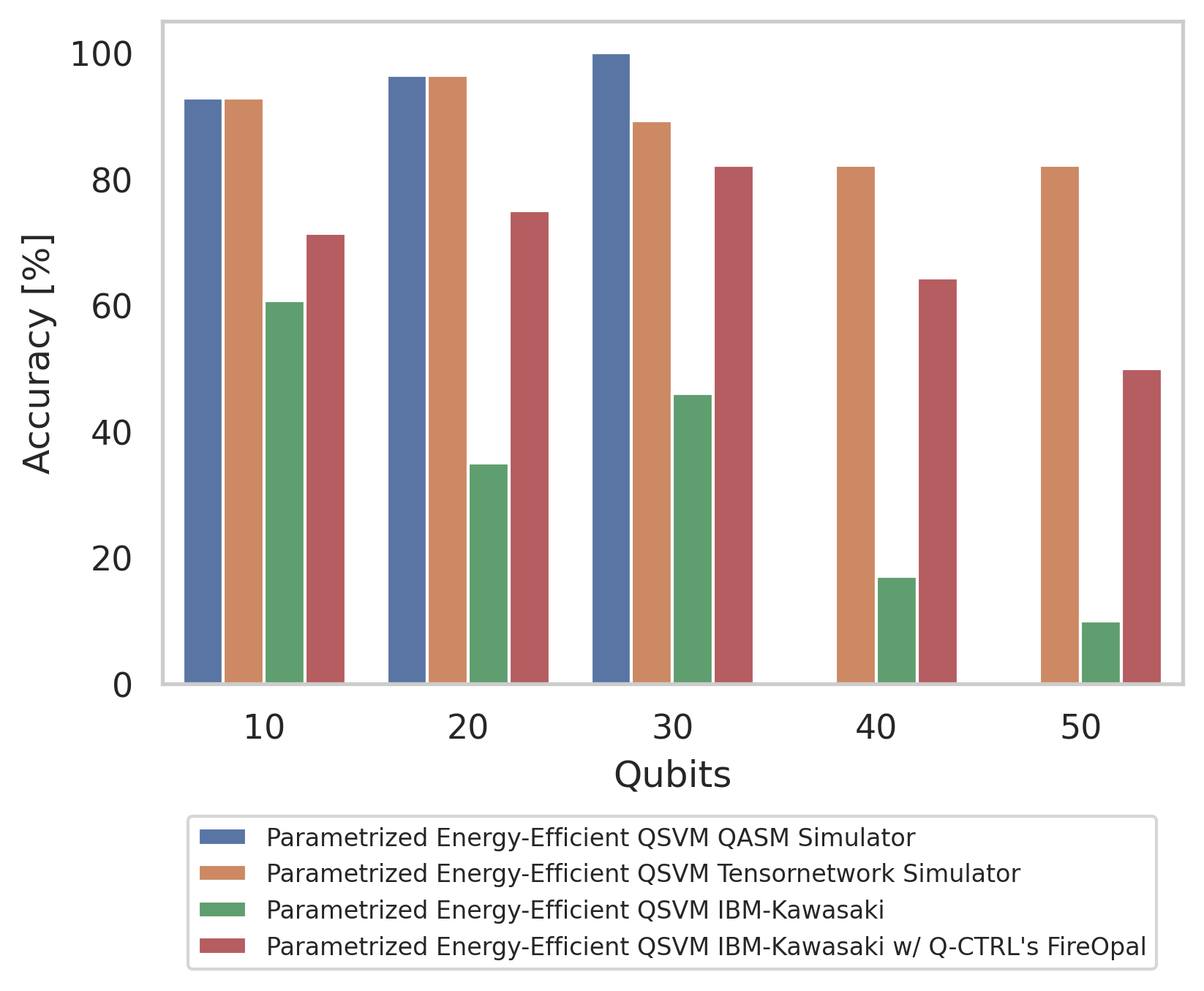}
\end{center}
\caption{{\small IBM's superconducting quantum computer evaluation. Compared to the classification performance of IBM-Kasawaki Standalone shown by the green bar, the performance using error suppression by Q-CTRL's FireOpal shown by the red bar is significantly improved, and the performance approaches the ideal value at 30 qubits.}}
\label{fig:graphhw}
\end{figure*}

\begin{figure*}[p]
\begin{minipage}{.98\textwidth}
\begin{center}
  \subfigure[IBM-Kawasaki w/ Q-CTRL's Fire Opal]{
   \includegraphics[width=120mm]{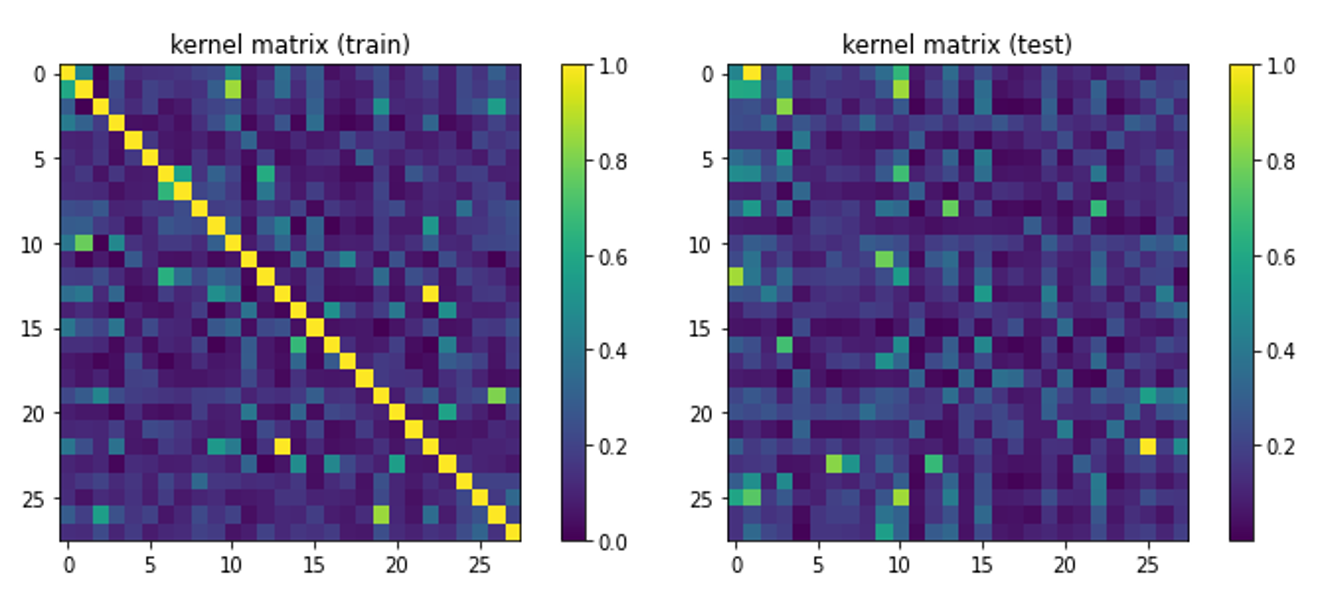}
  }
  
  \subfigure[IBM-Kawasaki Standalone]{
   \includegraphics[width=120mm]{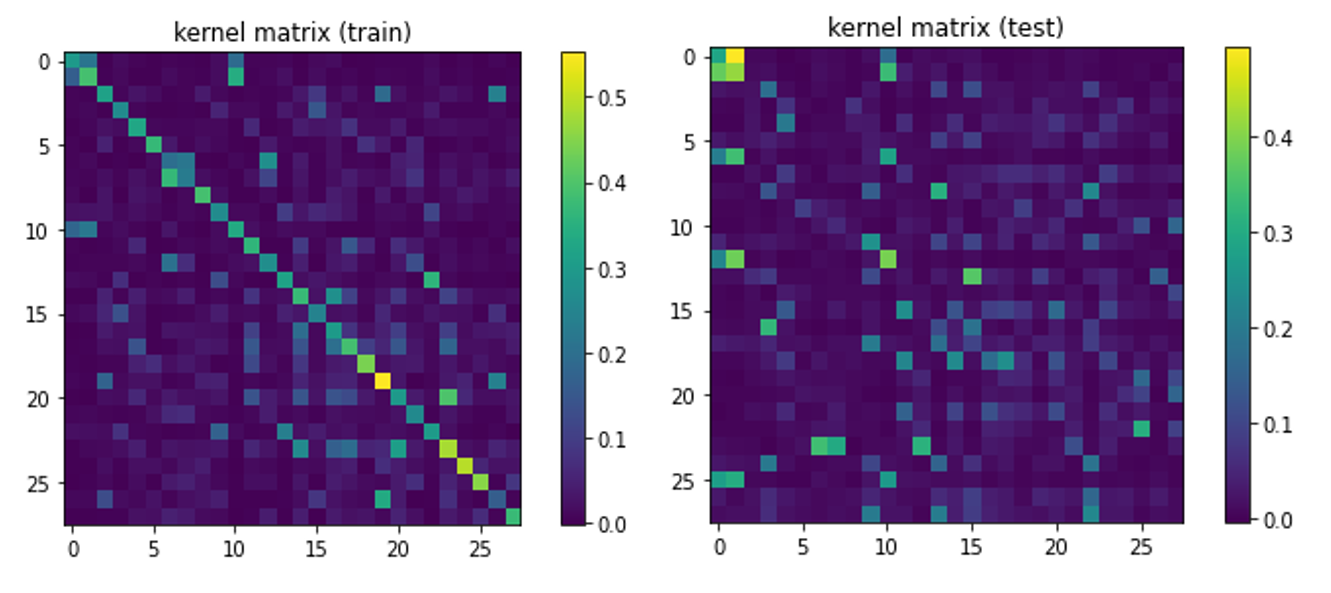}
  }
  \end{center}
\end{minipage}

\begin{minipage}{.95\textwidth}
  \begin{center}
  \subfigure[Inner product values of all data points of Test data]{
   \includegraphics[width=140mm]{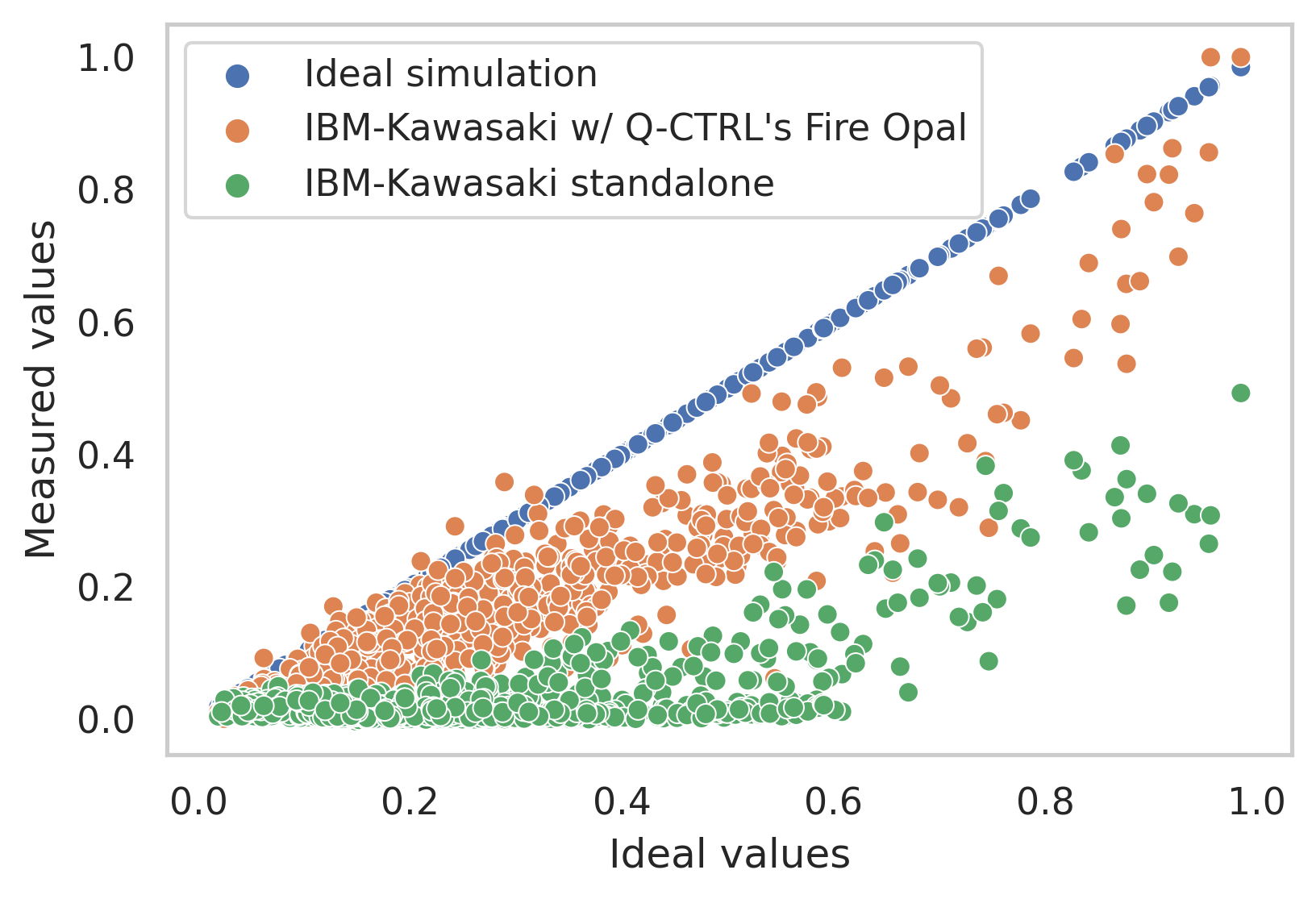}
  }
  \end{center}
\end{minipage}
  
\caption{{\small Gram matrix data comparison between IBM-Kawsaki w/ Q-CTRL's Fire Opal and IBM-Kawasaki Standalone, and distribution of calculated inner product values compared to each ideal value. (a) Gram matrix data at 30 qubits on IBM-Kawasaki after error suppression by Q-CTRL's Fire Opal. The left is the inner product between the training data, and the right is the inner product between the training and test data. At this time, the classification accuracy rate was 82\%. (b) Gram matrix data when applying the IBM-Kawasaki standard error suppression, the classification accuracy rate was 46\%. (c) Data comparing the inner product value on each Gram matrix with the ideal value determined by the simulator confirms that the performance of Q-CTRL's FireOpal in inner product calculation is uniform from 0 to 1.}} 
\label{fig:kernelmatrix}
\end{figure*}

Using the settings of the Parametrized Energy-Efficient quantum kernels confirmed in simulations, we conducted performance evaluation on IBM hardware.
{\mdseries Fig. \ref{fig:qpu}} shows the QPU architecture of 127-qubit IBM-Kawasaki\cite{bib:ibmq}, which is an IBM Quantum Eagle r3 processor, used in the experiment.

In this experiment, in addition to evaluating the tensor network simulator described in the previous section, we evaluated the state vector type simulator for IBM-qasm-simulator\cite{bib:qiskit}, with optimal parameters up to 30 qubits, the maximum number of qubits possible. 
Furthermore, we conducted evaluations on IBM-Kawasaki using the Sampler class in Qiskit Runtime with the standard error suppression algorithm, M3 (Matrix-free Measurement Mitigation), and circuit optimization level 3 applied. 
Regarding the explanation of the number of job executions in this experiment, the number of shots per job for the execution of quantum circuits in each inner product calculation was set to 4,000, and the number of quantum circuit calculations required to obtain a single kernel was 784 for both the training and test data, respectively.
Furthermore, we performed comparative evaluations with the effect of error suppression feature after applying Q-CTRL's Fire Opal.
The results are shown in {\mdseries Fig. \ref{fig:graphhw}}.

In this result, the performance of the tensor network simulator and IBM-qasm-simulator are used as our ideal values. 
The performance on IBM hardware is compared with the results obtained when using Q-CTRL's Fire Opal error suppression feature. 
Comparing the results of the simulators, IBM-qasm-simulator reaches 100\% the same quality output as simulation, at 30 qubits, whereas the performance of the tensor network deteriorates with the increase in the number of qubits. 
This deterioration is thought to be due to the expansion of approximation errors in the tensor network simulator and depletion of the number of data samples.

Regarding the results obtained from the hardware, performance declines beyond 30 qubits, which serves as the peak. 
Overall, the effect of error suppression by Q-CTRL's Fire Opal is significant, with the accuracy reaching 82\% at 30 qubits. 
Although it is not possible to compare with the results of IBM-qasm-simulator for more than 30 qubits, performance declines for some reason. 
This decline could be due to hardware noise or the increase in the number of qubits causing a deterioration in generalization performance due to the complexity of the dataset.

{\mdseries Fig. \ref{fig:kernelmatrix}} shows two types of data for Gram matrices 
in the above evaluation with 30 qubits calculated on a quantum computer, using IBM Standalone and after error suppression with Q-CTRL's Fire Opal, respectively, for training data pairs (left) and training data versus test data (right). 
The classification accuracy was 46\% for IBM Standalone and 82\% after error suppression with Q-CTRL's Fire Opal, respectively.
In the figure for training data pairs, the difference in accuracy due to inner product value calculations is noticeably distinct when looking at the diagonal elements. 
Additionally, the figure at the bottom compares measured inner-product value with the results from the IBM-qasm-simulator as an ideal value. 
Overall, it is clear that the results of inner product calculations in the Gram matrices using Q-CTRL's Fire Opal show performance closer to ideal values.

\section{Conclusion}

We propose an algorithm for a classification function in quantum machine learning and experimentally demonstrated its superiority over conventional quantum kernels by applying it to the use case of network service fault diagnosis.
By adjusting $\alpha$ parameter of the hyper-parameter related to the parametrization strength to the phase information in the quantum entanglement generation part, superior performance can be drawn out effectively. 

In terms of evaluating ideal values, evaluations were conducted using both state vector simulators and tensor network simulators, and hardware evaluations using IBM superconducting quantum computers were also performed. 

First, we conducted a performance evaluation of the proposed method using a tensor network simulator. 
Using the optimal settings, we evaluated the ideal values for both the tensor network simulator and the state vector simulator, and then assessed the performance using IBM hardware and error suppression with Q-CTRL's Fire Opal in comparison.

In the simulator experiment, we confirmed that the proposed method improves the classification performance compared to the conventional method. Also a divergence in performance related to approximation error factors was observed between the state vector type and the tensor network type.

In the hardware experiment with IBM-Kawasaki, although there is a degradation due to noise effects in absolute terms, by utilizing the error suppression feature of Q-CTRL's Fire Opal, it was possible to achieve classification performance close to ideal values even on NISQ machines.
This result suggests that quantum machine learning is approaching a level where it can be practically used.

Future studies will further investigate the causes of these degradation factors of tensor network simulator for the state vector simulator and explore further improvement methods for hardware performance.
There are also issues with scalability in learning, and it is necessary to consider methods for efficiently learning from larger data samples.
Furthermore, the relationship between the number of qubits and their representational power in relation to data complexity is also interesting.
Moreover, within this context, it is worth considering the challenge of verifying the superiority of quantum kernels over classical computing in the utility-scale.

Finally, we would like to restate the significance of this study that we experimented the implementation of quantum computers using data from a system that works a crucial role in our commercial operations, as a usecase for quantum machine learning. 
It was confirmed that quantum kernel learning has performance surpassing classical algorithms in simulations, and we acquired strategies for effectively utilizing it in circuit implementations. 
By employing these strategies, we were able to operate an IBM superconducting quantum computer in practice and verify that it can achieve robust performance with effective error suppression. 
Through this investigation, we believe that with just more improvement in hardware performance and modifications related to building scalable models, it is possible to implement quantum computers in commercial systems. 
We are grateful for the sense that the advantageous position of quantum computers is now within reach.

\section{Acknowledgement}
This work is partly supported by UTokyo Quantum Initiative and Q-CTRL.

\end{document}